\documentclass{article}
\usepackage{graphicx}

\usepackage{geometry}
\usepackage{amsthm, amsmath,amssymb, array}
\theoremstyle{definition}
\newtheorem{definition}{Definition}[section]
\newtheorem{corollary}{Corollary}[section]
\newtheorem{remark}{Remark}[section]
\theoremstyle{plain}
\newtheorem{theorem}{Theorem}[section]

\begin{document}

\large
\begin{center}
\textbf{Quasi-Feynman formulas for a Schroedinger equation\\ with a Hamiltonian equal to a finite sum of operators
}\end{center}

\begin{center}
Ivan\,D.~Remizov
\end{center}
\normalsize

\begin{center}
Bauman Moscow State Technical University\\
Rubtsovskaya nab., 2/18, office 1029. 105005 Moscow, Russia\\
Lobachevsky Nizhny Novgorod State University\\
Prospekt Gagarina, 23, Nizhny Novgorod, 603950, Russia
\end{center}

In this short communication I generalize the method of obtaining quasi-Feynman formulas described in my previous paper on that topic. The theorem presented allows to obtain the solution to the Cauchy problem for the Schr\"odinger equation with the Hamiltonian decomposed to a finite sum of operators. The concept of Chernoff tangency is used, and the solution is written in the form of a quasi-Feynman formula as before. Theorem proven is compared to known approximation theorems:  Trotter's, Chernoff's, Butko-Schilling-Smolyanov's. 

\section{Short introduction}

This preprint's aim is to present the statement and the proof of the theorem \ref{mainth}. The text is a continuation of \cite{Rem}, which is available for free from arXiv. In \cite{Rem} one can find wider introduction to the field, necessary background and the bibliography, as well as comments and examples to the abstract theorem proven. 

\begin{definition} A quasi-Feynman formula is a representation of a function in a form which includes multiple integrals of an infinitely increasing multiplicity. \end{definition}

Formulas (\ref{supadupaF1}) and (\ref{supadupaF2}) from theorem \ref{mainth} are examples 
of quasi-Feynman formulas if at least one of the families $(S_k(t))_{t\geq 0}$ consists of integral operators. The obtained formulas give the exact solution to the Cauchy problem for the Schr\"{o}dinger equation 

\begin{equation}\label{SE}
\left\{ \begin{array}{ll}
 i\psi'_t(t,x)=\mathcal{H}\psi(t,x);\quad t\in\mathbb{R}, x\in Q\\
 \psi(0,x)=\psi_0(x);\quad x\in Q\\
  \end{array} \right.
\end{equation}

where $\mathcal{H}$ is the Hamiltonian of the Schr\"odinger equation (\ref{SE}). The operator $\mathcal{H}$ is a self-adjoint operator in Hilbert space $\mathcal{F},$ and $\mathcal{F}$ is a space of wave functions, i.e. for each $t\in\mathbb{R}$ we have $\psi(t,\cdot)\in\mathcal{F}.$ Usually $\mathcal{F}$ is $L_2(Q),$ where $Q$ is a configuration space of the system described by (\ref{SE}). The vector $\psi_0\in\mathcal{F}$ is regarded as initial state of the system, and vector $\psi(t)=e^{-it\mathcal{H}}\psi_0$ is a state of the system at the time $t$.

In this paper I propose a method of evaluating $e^{-it\mathcal{H}}\psi_0$ for all $\psi_0\in\mathcal{F}$. It uses the decomposition  
$$-\mathcal{H}=L=a_1L_1+\dots+a_mL_m$$

and the existence of functions $S_k$ that are Chernoff-tangent to $L_k$.

\section{Short preliminaries}

Here only several definitions and basic theorems are listed, and only those that are used in the proof of the main theorem. Wider discussion and citations see in \cite{Rem}.
\vskip0.3cm 

This paper is about $C_0$-groups. The definition of a $C_0$-group is classical and is supposed to be known, but we state it here for the sake of good order. 

\begin{definition}Let $\mathcal{F}$ be a Banach space over the field $\mathbb{C}$. Let $\mathcal{L}(\mathcal{F})$ be a set of all bounded linear operators in $\mathcal{F}$. Suppose we have a mapping $V\colon [0,+\infty)\to \mathcal{L}(\mathcal{F}),$ i.e. $V(t)$ is a bounded linear operator $V(t)\colon \mathcal{F}\to \mathcal{F}$ for each $t\geq 0.$ The mapping $V$ is called a \textit{$C_0$-semigroup}, or \textit{a strongly continuous one-parameter semigroup} if it satisfies the following conditions: 

1) $V(0)$ is the identity operator $I$, i.e. $\forall \varphi\in \mathcal{F}: V(0)\varphi=\varphi;$ 

2) $V$ maps the addition of numbers in $[0,+\infty)$ into the composition of operators in $\mathcal{L}(\mathcal{F})$, i.e. $\forall t\geq 0,\forall s\geq 0: V(t+s)=V(t)\circ V(s),$ where for each $\varphi\in\mathcal{F}$ the notation $(A\circ B)(\varphi)=A(B(\varphi))$ is used;

3) $V$ is continuous with respect to the strong operator topology in $\mathcal{L}(\mathcal{F})$, i.e. $\forall \varphi\in \mathcal{F}$ function $t\longmapsto V(t)\varphi$ is continuous as a mapping $[0,+\infty)\to \mathcal{F}.$

The definition of a \textit{$C_0$-group} is obtained by the substitution of $[0,+\infty)$ by $\mathbb{R}$ in the paragraph above.
\end{definition}

\begin{remark}If $(V(t))_{t\geq 0}$ is a $C_0$-semigroup in Banach space $\mathcal{F}$, then the set $$\left\{\varphi\in \mathcal{F}: \exists \lim_{t\to +0}\frac{V(t)\varphi-\varphi}{t}\right\}\stackrel{denote}{=}Dom(L)$$ is dense in $\mathcal{F}$. The operator $L$ defined on the domain $Dom(L)$ by the equality $$L\varphi=\lim_{t\to +0}\frac{V(t)\varphi-\varphi}{t}$$ is called an infinitesimal generator (or just generator to make it shorter) of the $C_0$-semigroup $(V(t))_{t\geq 0}$. The generator is a closed linear operator that defines the $C_0$-semigroup uniquely, and the notation $V(t)=e^{tL}$ is used. If $L$ is a bounded operator and $Dom(L)=\mathcal{F}$ then $e^{tL}$ is indeed the exponent defined by the power series $e^{tL}=\sum_{k=0}^\infty\frac{t^kL^k}{k!}$ converging with respect to the norm topology in $\mathcal{L}(\mathcal{F})$. In most interesting cases the generator is an unbounded differential operator such as Laplacian $\Delta$. 
\end{remark}

The Stone theorem implies that $e^{-it\mathcal{H}}$ exists for each self-adjoint $\mathcal{H}$ and each $t\in\mathbb{R}$:

\begin{theorem} (\textsc{M.\,H.~Stone \cite{Stone}, 1932}) There is a one-to-one correspondence between the linear self-adjoint operators $H$ in Hilbert space $\mathcal{F}$ and the unitary strongly continuous groups $(W_t)_{t\in \mathbb{R}}$ of linear bounded operators in $\mathcal{F}$. This correspondence is the following: $iH$ is the generator of $(W_t)_{t\in \mathbb{R}}$, which is denoted as $W_t=e^{itH}.$ 
\end{theorem}

\begin{corollary}\label{cor} If $A$ is a linear self-adjoint operator in Hilbert space, then $\left\|e^{iA}\right\|=1.$ 
\end{corollary}

The following definition of Chernoff tangency (introduced in \cite{Rem}) plays the main role in what follows:

\begin{definition} Let $\mathcal{F}$ be a Banach space,
and $\mathcal{L}(\mathcal{F})$ be the space of all linear bounded operators in $\mathcal{F}$ endowed with the operator norm. Let $L\colon Dom(L)\to \mathcal{F}$ be a linear operator defined on $Dom(L)\subset\mathcal{F}$.

A function $G$ is said to be \textit{Chernoff-tangent} to the operator $L$ and the symbols $G\stackrel{Ch}{\asymp} L$ are written iff:

(CT1). $G$ is defined on $[0,+\infty)$, takes values in $\mathcal{L}(\mathcal{F})$ and $t\longmapsto G(t)f$ is continuous for every vector $f\in\mathcal{F}$. 

(CT2). $G(0)=I,$ i.e. for all $f\in\mathcal{F}$ we have $G(0)f=f.$ 

(CT3). There exists a dense subspace $\mathcal{D}\subset \mathcal{F}$ such that for every $f\in \mathcal{D}$ there exists
a limit $G'(0)f=\lim_{t\to 0}(G(t)f-f)/t$. 

(CT4). The operator  $(G'(0),\mathcal{D})$ has the closure $(L,Dom(L)).$ 

\end{definition}

The following Chernoff theorem allows to construct the $C_0$-semigroup in $\mathcal{F}$ having suitable family of linear bounded operators in $\mathcal{F}$. This family usually does not have a semigroup composition property but is pretty close to a $C_0$-semigroup in the sense described in the theorem below. For many $C_0$-semigroups such families $G$ are known or have been constructed in the past 15 years by the group of O.G.Smolyanov. We provide the wording of the Chernoff theorem based on the concept of the Chernoff tangency. 

\begin{theorem}\label{FormulaChernova}(\textsc{P.\,R.~Chernoff \cite{Chernoff}, 1968}) Let $\mathcal{F}$ be a Banach space,
and $\mathcal{L}(\mathcal{F})$ be the space of all linear bounded operators in $\mathcal{F}$ endowed with the operator norm. Let $L\colon Dom(L)\to \mathcal{F}$ be a linear operator defined on $Dom(L)\subset\mathcal{F}$. 

\textbf{Suppose} there is a function $G$ such that:

(E). There exists a strongly continuous semigroup $(e^{tL})_{t\geq 0}$ and its generator is $(L,Dom(L))$.

(CT). $G$ is Chernoff-tangent to $L.$

(N). There exists $\omega\in\mathbb{R}$ such that $\|G(t)\|\leq e^{\omega t}$ for all $t\geq 0$.

\textbf{Then} for every $f\in \mathcal{F}$ we have $(G(t/n))^nf\to e^{tL}f$ as $n\to \infty$, and the limit
is uniform with respect to $t$ from every segment $[0,t_0]$ for every fixed $t_0>0$.
\end{theorem}

From the Chernoff theorem one can infer this simple fact that will be used to obtain quasi-Feynman formulas:
\begin{corollary}\label{corboundop} If $\mathcal{F}$ is a Banach space, and $A\colon \mathcal{F}\to \mathcal{F}$ is a linear bounded operator, then 
$$e^A=\sum_{k=0}^\infty\frac{A^k}{k!}=\lim_{k\to\infty}\left(I+\frac{A}{k}\right)^k.$$
\end{corollary}

\begin{definition}\label{Cheq}Let $\mathcal{F}$ and $\mathcal{L}(\mathcal{F})$ be as before. Let us call two $\mathcal{L}(\mathcal{F})$-valued mappings $G_1$ and $G_2$ defined both on $[0,+\infty)$ (respectively, both on $\mathbb{R}$)  \textit{Chernoff-equivalent} (denoting this as $G_1\stackrel{Ch}{\sim} G_2$) if and only if  $G_1(0)=G_2(0)=I$ and for each $f\in\mathcal{F}$ and each~$T>0$
$$\lim_{n\to\infty}\sup_{\scriptsize\begin{array}{cc}
t\in [0,T]\\
(resp.\ t\in [-T,T]) \end{array}}\left\|\left(G_1\left(\frac{t}{n}\right)\right)^nf-\left(G_2\left(\frac{t}{n}\right)\right)^nf\right\|=0.$$
\end{definition} 

\begin{remark} There are several slightly different definitions of the Chernoff equivalence, I will just follow \cite{OSS} not going into details. The only thing I need from this definition is that if $G_1$ and $L$ satisfy all the conditions of the Chernoff theorem, then the mapping $G_1$ is Chernoff-equivalent to the mapping $G_2(t)=e^{tL}$. In other words, the limit of $(G_1(t/n))^n$ yields the $C_0$-semigroup $(e^{tL})_{t\geq 0}$ as $n$ tends to infinity.
\end{remark}

\begin{remark} With these definitions, the Chernoff-equivalence of $G$ to $e^{tL}$ follows via Chernoff theorem from the existence (E) of the $C_0$-(semi)group plus Chernoff-tangency (CT) plus the growth of the norm  bound (N). \end{remark}

\begin{remark} Denoting the Chernoff equivalence by $\stackrel{Ch}{\sim}$ and the Chernoff tangency by $\stackrel{Ch}{\asymp}$ one can compare the approximation theorems:
\end{remark}

\begin{center}
  \begin{tabular}{ | c | c  | c | c | }
    \hline
		
		\hline
\multicolumn{4}{ |c| }{How to obtain $\exp(tL)$} \\
\hline
		
    Case & We have & Then & History \\ \hline
		$L=A+B$ & $e^{tA}$, $e^{tB}$ & $e^{tL}=\lim\limits_{n\to\infty} \Big(e^{\frac{t}{n}A}\circ e^{\frac{t}{n}B}\Big)^n$  & Trotter \cite{Trotter}, 1958 \\ \hline

    General & $S(t)\stackrel{Ch}{\sim}e^{tL}$ & $e^{tL}=\lim\limits_{n\to\infty} \left(S\left(\frac{t}{n}\right)\right)^n$ & Chernoff \cite{Chernoff}, 1968 \\ \hline

$L=L_1+\dots+L_m$ &  $\forall k=1,2,\dots,m$ &  & Butko-Schilling\\ 

where $m\in\mathbb{N}$ is fixed  &$S_k(t)\stackrel{Ch}{\sim}e^{tL_k}$ & $e^{tL}=\lim\limits_{n\to\infty} \Big(S_1\left(\frac{t}{n}\right)\circ \dots\circ S_m\left(\frac{t}{n}\right)\Big)^n$ &-Smolyanov \cite{ButSS}, 2012\\ \hline

$L=iaH$      & $S(t)\stackrel{Ch}{\asymp}H$  &  $e^{tL}=\lim\limits_{n\to\infty} \left(R\left(\frac{t}{n}\right)\right)^n$ & Remizov \cite{Rem}, 2014 \\

where $H=H^*$, $a\in\mathbb{R}$ &  $S(t)^*=S(t)$                     & where $R(t)=\exp\big[ia(S(t)-I)\big]$ &\\ \hline

$L=iH,$ $H=H^*$  &  & & \\

$H=a_1H_1+\dots+a_m H_m$ & $\forall k=1,2,\dots,m$ &$e^{tL}=\lim\limits_{n\to\infty} \left(R\left(\frac{t}{n}\right)\right) ^n$ where  & Remizov, 2015 \\

where $m\in\mathbb{N}$ is fixed and & $S_k(t)\stackrel{Ch}{\asymp}H_k$,  & $R(t)=\exp\Big[i\sum_{k=1}^ma_k(S_k(t)-I)\Big]$  & (theorem \ref{mainth} below) \\ 

$a_k\in\mathbb{R}$ $\forall k=1,2,\dots,m$ &$S_k(t)^*=S_k(t)$ &  & \\ \hline

  \end{tabular}
\end{center}

\section{Main theorem}

\begin{theorem}\label{mainth}
Let $\mathcal{F}$ be a complex Hilbert space, and let $(L_k,Dom(L_k))_{k=1}^m$ be a finite set of linear (bounded or unbounded) operators $L_k\colon Dom(L_k)\to \mathcal{F},$ where $Dom(L_k)\subset\mathcal{F}$. Let $a_1,\dots,a_m$ be a finite set of non-zero (real or complex) numbers. 

\textbf{Suppose} that

1) For all $k=\overline{1,m}$ the function $S_k$ is Chernoff-tangent to the operator $L_k$.

2) In item 1) above one can take the same linear dense subspace $\mathcal{D}\subset\mathcal{F}$ in (CT3) and (CT4) for all $k=\overline{1,m}$. 

3) The closure $(L,Dom(L))$ of the operator $(a_1L_1+\dots+a_mL_m,\mathcal{D})$ is self-adjoint.

4) For each $t\geq 0$ the operator $a_1S_1(t)+\dots+a_mS_m(t)\stackrel{denote}{=}S(t)$ is self-adjoint.

5) The number $a=\sum_{k=1}^ma_k$ is real.

\textbf{Then} 

1. There exists a $C_0$-group $e^{itL}$ with the generator $iL$.

2. The following family of bounded linear operators is well-defined $$R(t)=\exp\left[i\sum_{k=1}^ma_k(S_k(|t|)-I)\, \mathrm{sign}(t)\right]=\exp\Big[i(S(|t|)-aI)\, \mathrm{sign}(t)\Big].$$ 

3. The family $(R(t))_{t\in\mathbb{R}}$ is Chernoff-equivalent to the group $\left(e^{itL}\right)_{t\in\mathbb{R}}.$ It means that for all $t\in\mathbb{R}$

\begin{equation}\label{supadupaF}e^{itL}f=\lim_{n\to\infty}(R(t/n))^nf\end{equation}

4. Denote $a=a_1+\dots+a_m,$ $a_{m+1}=-a$ and  $S_{m+1}(t)=I$ for all $t\geq 0$ for simpicity of the formulas below. Then   for each $f\in\mathcal{F}$ and each $t\in\mathbb{R}$ the following formulas hold:

\begin{equation} \label{supadupaF1}e^{itL}f=\lim_{n\to\infty}\lim_{j\to\infty}\sum_{p=0}^j \frac{(in\, \mathrm{sign}(t))^p}{p!} 
\sum_{k_1=1}^{m+1}\dots\sum_{k_p=1}^{m+1} a_{k_1}\dots a_{k_p} S_{k_1}(|t|/n)\circ \dots \circ S_{k_p}(|t|/n)f \end{equation}

\begin{equation}\label{supadupaF2}e^{itL}f=\lim_{n\to\infty}\lim_{p\to\infty} p!\,\sum_{d=0}^p \frac{(in\, \mathrm{sign}(t))^{d}}{p^dd!(p-d)!} \sum_{k_1=1}^{m+1}\dots\sum_{k_d=1}^{m+1} a_{k_1}\dots a_{k_d} S_{k_1}(|t|/n)\circ \dots \circ S_{k_d}(|t|/n)f\end{equation}

\end{theorem}
\begin{remark}
The conditions 4) and 5) of theorem \ref{mainth} trivially follow from the conditions 4') and 5') that are simpler in checking:

4') For each $t\geq 0$ and each $k=\overline{1,m}$ the operator $S_k(t)$ is self-adjoint.

5') Numbers $a_1, \dots, a_m$ real.
\end{remark}
\textbf{Proof of the theorem \ref{mainth}.} Item 1 follows from 3) and the Stone theorem. Item 2 follows from the fact that in the power index of the exponent for every fixed $t$ one finds linear bounded operators only thanks to 1), so the exponent can be defined as the power series  $e^A=\sum_{j=1}^\infty A^j/j!$ for $A= i\sum_{k=1}^ma_k(S_k(|t|)-I)\, \mathrm{sign}(t)$.  

To prove item 3 let us check the conditions of the Chernoff theorem for $R^+(t)=\exp\left[i\sum_{k=1}^ma_k(S_k(t)-I)\right]$ and $L$. The condition (E) follows from the Stone theorem: self-adjoint operator $(L,Dom(L))$ is the infinitesimal generator of the $C_0$-group $(e^{itL})_{t\in\mathbb{R}}$ and of the $C_0$-semigroup $(e^{itL})_{t\geq 0}$. 

Now let us check that $R^+$ is Chernoff-tangent to $(L,Dom(L))$.

CT1). The continuity of $t\longmapsto R^+(t)$ in the strong operator topology follows from the continuity of $t\longmapsto \sum_{k=1}^ma_k(S_k(t)-I)$ in the strong operator topology (recall (C1) for $S_k$) and the continuity of the exponent in the norm topology. So (CT1) for $R^+$ is completed. 

CT2). $R^+(0)=\exp\left[i\sum_{k=1}^ma_k(S_k(0)-I)\right]=\exp\left[i\sum_{k=1}^ma_k(I-I)\right]=e^0=I.$ 

CT3) i) Remember that (CT1) for $S_k$ says that for every $f\in \mathcal{F}$ the function $M_f\colon [0,+\infty)\ni t\longmapsto S_k(t)f\in\mathcal{F}$ is continuous. So by the Weierstrass extreme value theorem the set $M_f([0,1])\subset \mathcal{F}$ is compact and hence bounded for each $f\in\mathcal{F}$. This means that for each $f\in\mathcal{F}$ there exists a number $C_k(f)>0$ such that $\|S_k(t)f\|\leq C_k(f)$ for all $t\in[0,1]$. Next, by the Banach-Steinhaus uniform boundedness principle the family of linear bounded operators $(S_k(t))_{t\in[0,1]}$ is bounded collectively, i.e. there exists a number $C_k>0$ such that $\|S_k(t)\|<C_k$ for all $t\in[0,1]$. Finally, for all $t\in[0,1]$ and all $k=\overline{1,m}$ we have
$$\|S_k(t)f\|\leq C=\max(C_1,\dots,C_n).$$

ii) Suppose that linear operator $A\colon \mathcal{F}\to\mathcal{F}$ is bounded. Then $e^A=I+A+A^2\frac{1}{2!}+A^3\frac{1}{3!}+\dots=I+A+ \sum_{n=0}^\infty\frac{A^n}{(n+2)!}A^2\stackrel{denote}{=}I+A+ \Psi(A)A^2.$  One can see that $$\|\Psi(A)\|=\left\|\sum_{n=0}^\infty\frac{A^n}{(n+2)!}\right\|\leq\sum_{n=0}^\infty\frac{\|A\|^n}{(n+2)!}\leq\sum_{n=0}^\infty\frac{\|A\|^n}{n!}=e^{\|A\|}.$$

iii) Set $A=i\sum_{k=1}^ma_k(S_k(t)-I)$. Then for all $t\in[0,1]$ the estimates $\|A\|=\|i\sum_{k=1}^ma_k(S_k(t)-I)\|\leq \sum_{k=1}^m|a_k|(C+1)$ and $\Psi(i\sum_{k-1}^ma_k(S_k(t)-I))\leq \exp[\sum_{k=1}^m|a_k|(C+1)]\stackrel{denote}{=}M$ hold. So for all $t\in(0,1]$ we have
\begin{equation}\label{rpluseq}\frac{R^+(t)f-f}{t}= i\sum_{k=1}^ma_k\frac{S_k(t)f-f}{t} + h(t,f)\end{equation}
where
$$\|h(t,f)\| =\left\|i^2 \Psi\left(\sum_{k=1}^ma_k(S_k(t)-I)\right) \left(\sum_{j=1}^ma_j(S_j(t)-I)\right) \sum_{k=1}^ma_k\frac{S_k(t)f-f}{t}  \right\|\leq$$
$$\leq M \sum_{j=1}^m\sum_{k=1}^m|a_j||a_k|\left\|(S_j(t)-I) \frac{S_k(t)f-f}{t} \right\|\leq$$ 
/(CT3) for $S_k$ says that $\frac{S_k(t)f-f}{t}=L_kf + \alpha_k(t,f),$ and $\lim\limits_{t\to 0}\|\alpha_k(t,f)\|=0$/
$$\leq M \sum_{j=1}^m\sum_{k=1}^m|a_j|\cdot|a_k|\cdot\Big(\|(S_j(t)-I)(L_kf)\| + \|(S_j(t)-I)\alpha_k(t,f))\|\Big).$$
Now let $t\to 0.$ From (CT1) and (CT2) for $S_j$ we infer $\|(S_j(t)-I)(L_kf)\|\to~0.$ Relying on the fact that $\lim\limits_{t\to 0}\|\alpha_k(t,f)\|=0$ and on the estimate $\|(S_j(t)-I)\|\leq C+1$ we infer that $\|(S_j(t)-I)\alpha_k(t,f))\|\leq \|(S_j(t)-I)\|\cdot \|\alpha_k(t,f))\|\to 0.$ So one can proceed in (\ref{rpluseq}) to the limit as $t\to0$ and obtain 
$$\lim\limits_{t\to 0}\frac{R^+(t)f-f}{t}= iLf.$$

CT4) for $R^+$ is a part of the condition 3), which completes the proof of the fact that $R^+$ is Chernoff-tangent to $(L,Dom(L))$.

The (N) condition for $R^+$ with $\omega=0$ follows from the conditions 4), 5) and the corollary \ref{cor}. Indeed, the operator $S(t)-aI$ is self-adjoint so $\left\|\exp[i(S(t)-aI)]\right\|=1.$

All the conditions of the Chernoff theorem for $R^+$ are fulfilled, which implies that for all $t\geq 0$ and all $f\in\mathcal{F}$ the following formula holds: $$e^{itL}f=\lim_{n\to\infty}(R^+(t/n))^nf.$$ To prove (\ref{supadupaF}) substitute $t$ by $-t$, $a_k$ by $-a_k$ and apply the generation theorem for the groups from \cite{EN1} at p.79. 

Now let me prove (\ref{supadupaF1}) and (\ref{supadupaF2}). Formula (\ref{supadupaF}) says that $e^{itL}=\lim_{n\to\infty}\exp\big[in\sum_{k=1}^ma_k(S_k(|t/n|)-I)\, \mathrm{sign}(t)\big].$ To pass from this formula to (\ref{supadupaF1}) and (\ref{supadupaF2}) let me first calculate $\big[\sum_{k=1}^ma_k(S_k(|t/n|)-I)\big]^p$ for every $p\in\mathbb{N}$. Denote $a=a_1+\dots+a_m,$ $a_{m+1}=-a$ and  $S_{m+1}(t)=I$ for all $t\geq 0$ to get more regular form of the expression studied. Then $$\sum_{k=1}^ma_k(S_k(|t/n|)-I)=\sum_{k=1}^{m+1}a_kS_k(|t/n|)$$
and
\begin{equation}\label{helpful}\left(\sum_{k=1}^{m+1}a_kS_k(|t/n|)\right)^p=\sum_{k_1=1}^{m+1}\dots\sum_{k_p=1}^{m+1} a_{k_1}\dots a_{k_p} S_{k_1}(|t|/n)\circ \dots \circ S_{k_p}(|t|/n).
\end{equation}

Now set $A=in\sum_{k=1}^ma_k(S_k(|t/n|)-I)\, \mathrm{sign}(t)$ in formula $e^{A}=\sum_{p=1}^\infty A^p/p!$ and use (\ref{helpful}), which leads to (\ref{supadupaF1}).

Formula (\ref{supadupaF2}) is inferred from (\ref{supadupaF}) and the formula $e^{A}=\lim_{p\to\infty}(I+A/p)^p$ for the same operator $A$. First mention that
$$\left(I+\frac{A}{p}\right)^p= \left(\frac{
in\, \mathrm{sign}(t)}{p}\right)^p \left(\frac{p}{in\, \mathrm{sign}(t)}I + \sum_{k=1}^{m+1}a_kS_k(|t|/n)  \right)^p$$
then recall that $I$ commutes with all operators and apply Newton's binomial formula
$$\left(\frac{p}{in\, \mathrm{sign}(t)}I + \sum_{k=1}^{m+1}a_kS_k(|t|/n)  \right)^p=
\sum_{d=0}^p \frac{p!}{(p-d)!d!}\left(\frac{p}{in\, \mathrm{sign}(t)}I\right)^{p-d}\left(\sum_{k=1}^{m+1}a_kS_k(|t|/n)\right)^d.$$
Now mention that $$\left(\frac{in\, \mathrm{sign}(t)}{p}\right)^p\frac{p!}{(p-d)!d!} \left(\frac{p}{in\, \mathrm{sign}(t)}\right)^{p-d}=\frac{(in\, \mathrm{sign}(t))^{d}p!}{p^dd!(p-d)!}$$
then substitute $p$ by $d$ in (\ref{helpful}) and obtain
$$\left(I+\frac{A}{p}\right)^p=  p!\,\sum_{d=0}^p \frac{(in\, \mathrm{sign}(t))^{d}}{p^dd!(p-d)!}    \left(\sum_{k=1}^{m+1}a_kS_k(|t|/n)  \right)^d \quad \Box$$

\end{document}